\documentclass[10pt,twoside,twocolumn,english,showpacs,floatfix,
               aip,pof,longbibliography]{revtex4-1}
\usepackage[T1]{fontenc}
\usepackage[latin9]{inputenc}
\usepackage{amsmath}
\usepackage{graphicx}
\usepackage{amssymb}
\usepackage{babel}

\begin{document}

\title{Dynamics of Vesicles in shear and rotational flows: Modal Dynamics
and Phase Diagram}

\author{Norman J. Zabusky$^{*}$, Enrico Segre$^{\dagger}$,
 Julien Deschamps$^{\ddagger}$,
 Vasiliy Kantsler$^{\S}$ and Victor Steinberg$^{*}$ }

\affiliation{$^{*}$Physics of Complex Systems, Weizmann Institute of Science,
Israel\\
 $^{\dagger}$Physics Services, Weizmann Institute of Science,
Israel\\
 $^{\ddagger}$IRPHE, Aix-Marseille Université \& CNRS, Marseille,
France\\
 $^{\S}$DAMTP, Cambridge University, UK}
\begin{abstract}
Despite the recent upsurge of theoretical reduced models for vesicle
shape dynamics, comparisons with experiments have not been accomplished.
We review the implications of some of the recently proposed models
for vesicle dynamics, especially the Tumbling-Trembling domain regions
of the phase plane and show that they all fail to capture the essential
behavior of real vesicles for excess areas, $\Delta$, greater than
0.4. We emphasize new observations of shape harmonics and the role
of thermal fluctuations.
\end{abstract}


\maketitle

\section{Introduction}

In the last years there has been an upsurge of interest in the dynamical
response of micro-objects to low Reynolds number shear and elongation
flows. A large number of researchers have dealt with vesicles, including
the groups of Steinberg
(experiments)\cite{KantslerPRL2005,KantslerPRL2006,JulienShear,JulienMill},
Gompper (simulations)\cite{NoguchGompperShear,NoguchiGompperPNAS,NoguchiGompperFluct,Messlinger},
Lebedev (theory)\cite{LTVnjp,LTVpre} and Misbah
(theory and simulations)\cite{Danker,Farutin,Kaoui,Vlakhovska}.
Others have studied,
along similar lines, their next-of kin, capsules, like Seifert, Finken
and Kessler (theory and simulations)
\cite{KesslerFinkenSeifert,KesslerFinkenSeifert-swinging},
Skotheim and Secomb (theory and simulations) \cite{SkotheimSecomb}
and Bagchi and Kalluri (simulations)\cite{Bagchi}. A common theme
is the behavior of the microscopic object at the verge of dynamic
regime transition. The different investigators have discussed an extended
phenomenology of intermediate and purportedly peculiar motion regimes,
supporting their findings by the use of the one or the other model,
or numerical experiment.

For the simulations, there are two modes of investigation: one is
direct numerical simulation (DNS) in
two\cite{Messlinger,Biros2D,FinkenLamura,BirosAxi3D}
or three\cite{NoguchiGompperFluct} dimensions, which requires
computational sophistication; the other involves the derivation of
reduced dynamical models, usually in a \emph{perturbative} framework
for \emph{nearly-spherical} objects, $\Delta<<1$, which yields a
number of coupled nonlinear ODEs. The excess area of the vesicle
 $\Delta=(A/r_{0}^{2})-4\pi$
is assumed $\ll1$, where $A$ is the vesicle surface area and $r_{0}$
its characteristic radius, obtained from the vesicle volume
 $V=(4\pi/3)r_{0}^{3}$).
Most of the experimental data\cite{KantslerPRL2005,KantslerPRL2006,JulienShear,JulienMill}
for Tank Treading (TT) is in the range $0.05<\Delta<2$ and for Trembling
(TR) (called also vacillating-breathing, VB\cite{Misbah2006}, and
swinging\cite{NoguchiGompperSwinging}) is within $0.45<\Delta<2$.

The system of ODEs is then studied as a dynamical system, either analytically
or numerically, and its phase space properties are listed. Proposed
systems of this sort have included two\cite{LTVnjp,LTVpre,Danker,Kaoui},
three\cite{NoguchiSynchronized}, and recently 14 nonlinear ODE\cite{Farutin}.
A recent review was presented by \citet{Vlakhovska} which refers to
the latest work of Misbah and colleagues\cite{Farutin} and
includes vesicle results. Three important and inter-related issues,
often neglected by most of the theoretical and numerical work, are: a)
the effect of thermal fluctuations (because vesicles are small and
the bending energy of the membrane is comparable to the thermal energy);
b) the applicability of perturbative results, to the more readily
obtained excess areas, $\Delta>0.4$; and c) the role of odd modes
of the contour shape.

Following the literature, we survey three possible regimes of motion,
namely (TT), (TR) and tumbling (TU). We present here as our benchmark
an expanded data set including recent experimental work\cite{JulienMill},
with more attention to the dynamics of mode interactions. Experiments
of longer duration and higher resolution are ongoing and will allow
us to refine this benchmark. We also present a new analysis of our
earlier published TT data\cite{KantslerPRL2005,KantslerPRL2006},
demonstrating the success of the scaling with respect to the dimensionless
parameter $\Lambda$ (defined in eq.~(\ref{eq:SLambdaTau}) below).

The layout of the paper is as follows: in section \ref{sec:Dynamical-models}
we critically review the dynamical models for vesicle shapes proposed
in literature, which suggest useful non-dimensional parameters and
provide a basis for laws to scale the experimental data. In section
\ref{sec:ExperimentAnalysis} we present a new analysis of early and
recently published data on vesicles and properties of the three dynamical
regimes observed: in subsection \ref{sub:methods} we recall the experimental
setup and the methodology used to analyze vesicle contours, and in
the following \ref{sub:Tank-Treading}, \ref{sub:ExperimentModes}
and \ref{sub:periods} we review experimental findings concerning
vesicles in their regimes. In section \ref{sec:Phase-diagrams} we
consider the broader problem of the phase diagram of all regimes of
motion. Section \ref{sec:Conclusions} presents our conclusions.

\section{Dynamical models for vesicle inclination angle and shape evolution
         \label{sec:Dynamical-models}}

\subsection{Models without noise\label{sub:Models-without-noise}}

Among the reduced dynamical models mentioned in the introduction,
we refer in detail to three which are derived from first principles
for a near spherical vesicle, $\Delta\ll1$. A perturbation expansion
in powers of $\sqrt{\Delta}$ had already been given by \citet{Seifert}
for Helfrich membrane, which is represented by a series of spherical
harmonics (modes). If harmonics greater than two, thermal noise and
nonlinear terms above the third order in a free energy expansion are
neglected, one obtains a system of two coupled ODE's: namely the system of
Lebedeev, Turitsin and Vergeles\cite{LTVnjp,LTVpre}
(LTV), that of \citet{Danker} (DBPVM),
and its later variant\cite{Kaoui}
(KFM). We concentrate on these models because of their central role
in the recent discussion about regime transitions, because they do
not contain ad hoc fitted parameters (as the three equation system
of \citet{NoguchiSynchronized} does), and because they can be easily
simulated.

The ODE's describe the vesicle in terms of two dynamical variables
$\Theta$ and $\psi$, associated with the shape and tilt, respectively.
To avoid confusion we preserve the notation of most of the original
papers, despite that the tilt $\psi$ will be the quantity to be compared
to the experimental inclination angle $\theta$, and not $\Theta$.
The different regimes of vesicle motion are sought among the attractors
of the resulting dynamical system: for instance a fixed point with
positive $\psi$ is identified with tank treading (TT), a limit cycle
spanning the whole range $[-\pi,\pi]$ for $\psi$ is identified with
tumbling (TU), and a limit cycle spanning just the part of the range
for $\psi$ is identified with trembling (TR). Non-dimensional parameter
choices determine the phase portrait of the system, the structural
stability of phase space trajectories, and other issues, which are
of concern in this paper. Domains in the parameter space leading to
the one or the other dynamical regime, and their boundary lines, are
referred as to the phase diagram, discussed further in section
\ref{sec:Phase-diagrams}.

The equations of DBPVM read \begin{eqnarray}
\tau\partial_{t}\psi & = & \quad\frac{S}{2}\left[\frac{\cos2\psi}{\cos\Theta}\left(1+\sqrt{\Delta}\Lambda_{2}\sin\Theta\right)-\Lambda\right]\nonumber \\
\tau\partial_{t}\Theta & = & \negthickspace-S\left[\sin\Theta-\sqrt{\Delta}\Lambda_{1}\left(\cos4\Theta+\cos2\Theta\right)+\right.\nonumber \\
 &  & \qquad\left.-\sqrt{\Delta}\Lambda_{2}\cos2\Theta\right]\sin2\psi+\cos3\Theta\label{eq:DBPVM}\end{eqnarray}
 and involve the non-dimensional parameters: \begin{equation}
S=\frac{14\pi\eta r_{0}^{3}}{3\sqrt{3}\kappa}\frac{s}{\Delta},\enskip\Lambda=\frac{\left(23\lambda+32\right)}{8\sqrt{30\pi}}\frac{\omega}{s}\sqrt{\Delta},\enskip\tau=\frac{S\Lambda\sqrt{\Delta}}{2\omega},\label{eq:SLambdaTau}\end{equation}
 \[
\Lambda_{1}=\frac{\sqrt{10}}{28\sqrt{\pi}}\left(\frac{49\lambda+136}{23\lambda+32}\right),\quad\Lambda_{2}=\frac{10\sqrt{10}}{7\sqrt{\pi}}\left(\frac{\lambda-2}{23\lambda+32}\right),\]
 where $\lambda=\eta_{in}/\eta_{out}$ is the viscosity contrast,
$s$ and $\omega$ are respectively the strain rate and the vorticity
of the ambient flow, and $\kappa$ is the vesicle membrane bending
rigidity modulus. We note that while $\Lambda$ depends on both $\Delta$
and $\lambda$, $\Lambda_{1}$ and $\Lambda_{2}$ depend on $\lambda$
alone. In the original form\cite{Danker} these parameters are expressed
using the dimensionless shear rate $\chi$ (also called the capillary
number, $Ca$), i.e. \begin{equation}
\chi=Ca=\frac{\eta r_{0}^{3}\dot{\gamma}}{\kappa},\enskip S=\frac{7\pi}{3\sqrt{3}}\frac{\chi}{\Delta},\enskip\Lambda=\frac{23\lambda+32}{8\sqrt{30\pi}}\sqrt{\Delta},\label{eq:CaSLambdaShear}\end{equation}
 \[
\tau=\frac{7\sqrt{\pi}(23\lambda+32)}{72\sqrt{10}}\frac{\chi}{\dot{\gamma}}\ .\]
 We remark that this form refers to a pure shear flow, for which $s=\omega=\dot{\gamma}/2$.
The set of parameters $\{\chi,\lambda,S,\Lambda,\Delta\}$ is in fact
redundant, at least in pure shear; we include here all the definitions
for reference, as different sets are used in the original papers.

The DBPVM model differs from the LTV in that it introduced {}``higher-order''
terms, with additional parameters $\Lambda_{1}$ and $\Lambda_{2}\ll\Lambda_{1}$
which are multiplied by the small parameter $\sqrt{\Delta}$ in equations
(\ref{eq:DBPVM}). Further work by the same group\cite{Kaoui} employed
a variant of this model (KFM), in which only $\Lambda_{1}$ is present.
The simpler LTV model, in turn, can be obtained from DBPVM just by
setting $\Lambda_{1}=0$, $\Lambda_{2}=0$.

A further zero-temperature, 14-ODE dynamical model was introduced
by \citet{Farutin}, including the second and fourth spherical harmonics
and neglecting all others. This model purports to be most realistic,
although its results also disagree with the experiments presented
here. The merit of this model seems to be its good agreement with
unpublished 3D numerical simulations. Unfortunately, no details
about either the analytical or the computational model are given;
odd modes are absent since they would not be excited on an initially
ellipsoidal vesicle and the applied shear (both are described by j=2 modes),
in contradiction
to experimental observations\cite{KantslerPRL2006,JulienShear,JulienMill}.
As shown in section \ref{sub:ExperimentModes}, this is not the case,
especially in the TU and TR regimes. Also the recent direct numerical
simulations with thermal noise\cite{Messlinger} mentioned below in section
\ref{sub:Simulations-with-noise}
show TR states that are greatly distorted and contain odd harmonics.

The model used by \citet{NoguchiGompperFluct,NoguchiGompperSwinging},
which also results in a system of two ODE's, can also be quoted in
this respect, though in part phenomenological. This model expresses
the dynamics of the vesicle inclination angle $\theta$ and of a shape
parameter (asphericity) $\alpha$ by means of terms which have partially
a theoretical justification and are partially the result of numerical
evaluations on ellipsoidal shells. The inclination angle $\theta$
and the asphericity $\alpha$ of the object are coupled, and ad-hoc
fit of the free energy function dependent on $\alpha$ is employed.

\begin{table*}
\noindent \begin{centering}
\begin{tabular}{|c|c|c|c|c||c|c|}
\hline
Paper  & model type  & \textbf{$\Delta$}  & $S$  & $\chi$  & \textbf{ $\Lambda_{TR}$}  & $\lambda_{TR}$ \tabularnewline
\hline
\hline
Farutin et al.\ \cite{Farutin}  & 14 ODE, $q=2,4$  & \textbf{$0.43$} & $98.42$ & 10  & \textbf{$1.83\div2.6$}  & $8\div12$\tabularnewline
\hline
Vlahovska et al.\cite{Vlakhovska}  & 14 ODE, $q=2,4$  & \textbf{$0.43$} & $12$ & 1.22  & \textbf{$2\div2.2$}  & $8.91\div9.94$\tabularnewline
\cline{3-7}
  &  & \textbf{$1.0$} & $12$ & 1.22  & $2.38\div2.9$  & $8.91\div9.94$\tabularnewline
\hline
Noguchi and & 2 ODE, & \textbf{$0.44$} & $12$ & 1.24 & $1.82\div1.95$ & $7.86\div8.5$\tabularnewline
\cline{3-7}
Gompper\cite{NoguchiGompperSwinging,NoguchiOscillatory} & phenomenological & \textbf{$0.91$} & $36.9$ & 8 & $2.05\div2.2$ & $5.86\div6.36$\tabularnewline
\cline{3-7}
 &  & \textbf{$1.44$} & $23.5$ & 8 & $2.25\div2.4$ & $4.96\div5.37$\tabularnewline
\hline
\begin{tabular}{c}
Danker et al.\ \cite{Danker}\tabularnewline
(DBPVM)\tabularnewline
\end{tabular} & \begin{tabular}{c}
2 ODE, $q=2$\tabularnewline
$\Lambda_{1}\neq\Lambda_{2}\neq0$\tabularnewline
\end{tabular} & \multicolumn{1}{c|}{\textbf{$0.43$}} & $98.42$ & 10  & $1.43\div1.56$  & $5.97\div24.01$\tabularnewline
\hline
\multicolumn{1}{|c|}{Kaoui et al.\ \cite{Kaoui}} & 2 ODE, $q=2,$  & \multicolumn{1}{c|}{\textbf{$0.2$}} & $211.61$ & 10  & $1.42\div1.56$  & $9.33\div10.39$\tabularnewline
\cline{3-7}
\multicolumn{1}{|c|}{(KFM)} & $\Lambda_{1}\neq0$, $\Lambda_{2}\equiv0$  & \multicolumn{1}{c|}{\textbf{$0.43$}} & $98.42$ & 10  & $1.42\div1.57$  & $5.92\div6.70$\tabularnewline
\cline{3-7}
\multicolumn{1}{|c|}{} &  & \multicolumn{1}{c|}{\textbf{$1.0$}} & $42.32$ & 10  & $1.42\div1.615$  & $3.51\div4.08$\tabularnewline
\hline
\multicolumn{1}{|c|}{LTV \cite{LTVnjp}} & \begin{tabular}{c}
\multicolumn{1}{c}{2 ODE, $q=2$}\tabularnewline
$\Lambda_{1}\equiv\Lambda_{2}\equiv0$\tabularnewline
\end{tabular} & \multicolumn{1}{c|}{\textbf{$0.43$}} & $98.42$ & \multicolumn{1}{c||}{10} & $1.41\div1.5$  & $5.87\div6.33$\tabularnewline
\hline
\end{tabular}

\par\end{centering}

\caption{\label{tab:Comparison-of-ranges}Comparison of ranges of parameters
$\Lambda$ and $\lambda$ where trembling is observed, according to
different models. While we relied on information contained in the
paper referenced for the first four cases, we numerically simulated
the dynamical system for the three two-ODE models, DBPVM, KFM and
LTV. $q$ denotes the harmonic modes included in the dynamical model.}

\end{table*}

To benchmark the different models, aside of results available in the
literature, we compared the predictions about the range of $\Lambda$
or $\lambda$ for which the TR regime should be observed, at fixed
values of $\Delta$. We choose $\Lambda$ or $\lambda$ because all
proposed models predict TR in a stripe of the respective parameter
space, with a weaker dependence on the second parameter, which is
$S$ or $\chi$. Table \ref{tab:Comparison-of-ranges} presents a
summary of values obtained by different models - some as reported
by the respective papers, some reproduced by our numerical solutions
of these models. We confirmed the results for DBPVM, KFM and LTV in
a direct way, integrating the equations in MATLAB using the \texttt{ode45()}
integrator. We produced phase trajectories for selected values of
the control parameters, and noted the ranges of values of the parameters
at which different regimes of motion occur. We did not attempt to
reproduce the results of the 14 equation system of \citet{Farutin}
(as insufficient details are given), and we report here data extracted
from their original paper. The data for the model by
\citet{NoguchiGompperSwinging} was derived from their Fig.~4b,
(equivalent to Fig.~2a of \citet{NoguchiOscillatory}).
Fig.\ 2a of \citet{NoguchiOscillatory}
is in fact a phase diagram for $V^{*}=0.95,\,0.9,\,0.85$ 
(or $\Delta=0.44,\,0.91,\,1.44$,
$\Delta=4\pi\left[\left(1/V^{*}\right)^{2/3}-1\right]$) which is
compared with experimental data further below in Figure
\ref{fig:phase-diagrams-Noguchi}.
We based our comparison wherever possible on the parameter values
$\Delta=0.43$ and $\chi=10$, because these are recurring values
in the papers we refer to. From \citet{Vlakhovska},
we extracted the
range for $\Lambda$ from their Fig.\ 6, where $S\leq12$, corresponding
to just $\chi=1.22$. In all models the boundaries of TR motion are
anyway seen to depend little on $\chi$ at large $S$.

We can draw several conclusions from this comparison. The first and
main one is the disagreement in the number of independent non-dimensional
parameters required by each model: \emph{all, except LTV, require
more than two}, in contrast with what is observed in the experiment.
The second conclusion is more quantitative: none of the models really
predicts the observed parameter range for trembling motion (section
\ref{sec:Phase-diagrams}). If we look at $\Lambda$, in particular,
the differences in the predictions between DBPVM, KFM at different
$\Delta$, and LTV at any $\Delta$ turn out to be negligible, but
all three models predict a narrower range of $\Lambda$ for TR than
observed. Also, the model of \citet{Farutin} gives a prediction
disagreeing with every other model as well with
the experiment. Finally, we remark that the TR regime is inadequately
described in these models adopting a framework of the second and fourth
harmonics without thermal noise, as a 
\textquotedbl{}vacillating-breathing\textquotedbl{}
mode. As was pointed in our previous work\cite{JulienShear,JulienMill},
thermal noise
and third harmonics are crucial for understanding the dynamics of
the TR state.

\subsection{Simulations with noise\label{sub:Simulations-with-noise}}

\citet{Messlinger} present the results of two
dimensional simulations of vesicles based on multi-particle collision
dynamics, or MPC numerical algorithms. This is a discrete particle
method akin to dissipative particle dynamics and implicitly includes
fluctuations due to thermal motion. It has been developed by Gompper,
Noguchi and colleagues and applied to simulation of vesicles in two
and three dimensions. Moreover, \citet{NoguchiOscillatory},
in his comprehensive study of vesicles forced with oscillatory shear
flows, also includes MPC simulations with constant shear. His Fig.\ 1
presents selected shapes during TT, TR, and TU.

It is remarkable that the simulation in 2D, with noise, of
Messlinger et al.\cite{Messlinger}
captures the essence of what we observe experimentally. The vesicle
shapes, Fig.\ 3 (and the supplementary animation) of their paper
show a {}``swinging'' or {}``trembling'' motion which exhibits
concave regions of negative curvature in the contour and clearly includes
higher odd and even harmonics of the radial displacement. Such shapes
are similar to those we show in Figures \ref{fig:Imagery} and \ref{fig:spectra-and-contours}
below. Also the tumbling motion shown by \citet{Messlinger} contains higher odd
and even harmonics, but unfortunately no power spectra which could
be compared with our experiments (section \ref{sub:ExperimentModes}).

\citet{Messlinger} also compare the MPC simulations with the
two ODE model for the vesicle shape of Noguchi and Gompper, mentioned
in section \ref{sub:Models-without-noise}, which is studied with
the addition of noise forcing terms to both equations. In Fig.\ 5,
\emph{ibidem}, the MPC results are compared to phase space trajectories
of the 2-ODE model, with and without forcing (left column MPC, right
column, 2-ODE). Trajectories of the $\alpha$ vs.\ $\theta$ are
plotted for various parameter values, and it is apparent that, in
all cases but the first, swinging and tumbling occur \emph{intermittently},
i.e., for a large parameter range the dynamics is strongly dependent
on the noise forcing. This behavior has also been remarked in the
experiments, and leads to ambiguous classification of our TU-TR transition.
Unfortunately, a complete phase diagram, comparable with others, is
not included - Fig.\ 2 of \citet{Messlinger} has a sample of only
four points.

Noguchi and Gompper also discuss in their earlier
papers\cite{NoguchiGompperFluct,NoguchiGompperSwinging}
the addition of stochastic forcing terms to their system of two ODEs,
which are identical in form to those in \citet{Messlinger}. However,
in \citet{NoguchiGompperFluct} the effect of forcing is assessed only
with respect to TT, while all the phase diagrams shown in their other
papers are obtained in absence of stochastic forcing.

\section{Experimental analysis of vesicle dynamics in shear and general flows
         \label{sec:ExperimentAnalysis}}

In the following, we review experimental findings about vesicles in
either of the three dynamical regimes mentioned, with particular attention
to scaling laws and unifying parameters suggested by the theories
discussed in section \ref{sec:Dynamical-models}. We present a quantitative
comparison of the old\cite{KantslerPRL2005,KantslerPRL2006} and
new\cite{JulienShear,JulienMill} data on the inclination angles
of vesicles in TT motion, which was obtained by two
different approaches and analyzed differently in the two sets of
papers, and comment on the issue of the transition boundary from TT
to TU and TT to TR regimes as function of $\Delta$ and $\lambda$,
in old and new sets of the data. Then we present a new spectral analysis
of shapes, in particular for long time series in TR regimes. We also
relate to the recurrent motion observed in TR and TU, and comment
about the observed time periodicity.

\subsection{Experimental techniques, analysis,
    and definition of regime transition lines\label{sub:methods}}

Fluorescent vesicles with prescribed viscosity contrast $\lambda$
are prepared and followed when immersed in either planar ambient flow
with controlled strain $s$ and vorticity $\omega$, or shear flow
with controlled shear rate $\dot{\gamma}=2s$ in plane Couette and
channel flow configurations. Many different vesicles are imaged in
isolation, each for several values of either $s$, $\omega$ in plane
linear flow, or of $\dot{\gamma}$ in shear flow. The experiments
have been described previously, and we refer to our
papers\cite{KantslerPRL2005,KantslerPRL2006,JulienShear,JulienMill}
for all details of the procedure. We remind that vesicles with
$1\le\lambda\le9.19$
were studied in a shear flow\cite{KantslerPRL2005,KantslerPRL2006,JulienShear},
while only vesicles with $\lambda=1$ were studied in general linear
flow in \citet{JulienMill}. The bending rigidity modulus of the vesicles
was estimated to be $\kappa\simeq 25 k_B T$ in all the experiments.

\begin{figure*}
\noindent \begin{centering}
\includegraphics[width=0.7\textwidth]{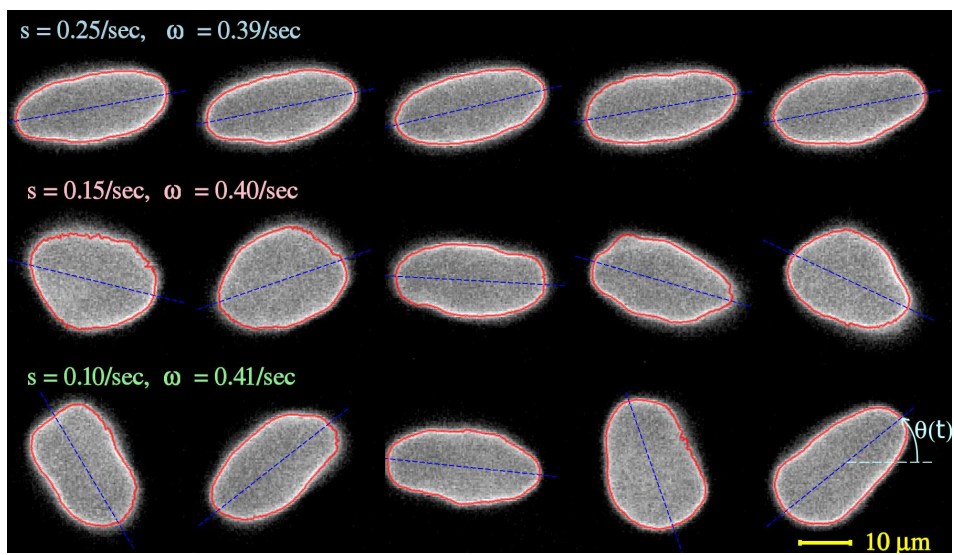}
\par\end{centering}

\caption{\label{fig:Imagery}Imagery of a single vesicle with $\lambda=1$
and $\Delta=1.13$ driven into all regimes of motion. Snapshots are
recorded every 1-2 sec. The red line (color online) is the reconstructed
sectional contour, further analyzed for harmonics, and the blue dashed
line is the major axis of the elliptical fit to the contour, according
to whose motion the regime is classified automatically. The first
row shows the vesicle in tank treading (note that the shape is almost
elliptical, though random fluctuations are appreciable); the second
row shows the vesicle during trembling (observe the remarkable changes
of shape, with an occasional trilobed shape, indicative of a significant
third harmonic). The last row shows the vesicle during tumbling.
Here the definition of $\theta$ is presented.}

\end{figure*}

Our experimental technique, described in \citet{JulienShear}, is capable
of determining $r_{0}$ and $\Delta$ from the full reconstruction
of the vesicle in three dimensions, but follows the vesicle motion
only through the imaging of sectional cuts. Care is taken to maintain
focus and obtain the largest (equatorial) section of the vesicle under
examination. We maintain that the variations in contour shape observed
are indicative of the dynamics taking part in three dimensions. In
Figure \ref{fig:Imagery} we show a representative example of an individual
vesicle (defined by $\Delta$ and $\lambda$) that changes its regime
of motion as the ambient general linear flow is changed by variation
of $\omega/s$.

We analyze the contours of the vesicle images, reconstructed at sub-pixel
accuracy by an ad hoc fitting algorithm which locates the brighter
edge of the vesicle. Such contours are then fitted by the ellipse
possessing the same tensor of inertia of the vesicle sectional area,
to define the major axis and the dominant orientation
$-\pi/2\le\theta(t)\le\pi/2$.
Two different approaches to determine transition thresholds between
TT and either TU or TR regimes were used in our
earlier\cite{KantslerPRL2005,KantslerPRL2006}
and later\cite{JulienShear,JulienMill} papers. In the
latter\cite{JulienShear,JulienMill},
the regime of motion of the vesicle, at time extents during which
flow parameters are kept constant, is classified automatically according
to an empirical criterion:
\begin{itemize}
\item if $\sqrt{\left\langle \left(\theta(t)-\left\langle \theta\right\rangle \right)^{2}\right\rangle }>\pi/5$,
then the vesicle regime is tumbling (TU);
\item if $\left\langle \theta\right\rangle >\sqrt{\left\langle \left(\theta(t)-\left\langle \theta\right\rangle \right)^{2}\right\rangle }$,
then the vesicle regime is tank treading (TT);
\item otherwise, the regime is classified as trembling (TR).
\end{itemize}
in which $\left\langle \cdot\right\rangle $ denotes the time average.
This criterion, based solely on the mean and rms fluctuation values
of the inclination angle and somewhat arbitrary thresholds, proved
itself simple and robust for isolated vesicles, well defined in shape.
It works effectively in presence of noisy data, and avoids the need
of phase-unwrapping the inclination angle to resolve TU.

The approach to the data undertaken in the early papers\cite{KantslerPRL2005,KantslerPRL2006}
is somewhat different, and needs to be elucidated before comparing
the various data sets. The early experiments were conducted to investigate
the TT dynamics, namely the dependence of the inclination angle $\theta$
on $\Delta$ and $\lambda$. In order to reduce the scatter in the
data due to thermal noise, in particular at smaller values of $\lambda$,
measurements were averaged over large ensembles of more than 500 vesicles
of the same value of $\lambda$ (see, for example, typical data in
Fig.~\ref{fig:Vasiliy1}). At that time the value of $\lambda$ could
only be inferred from the preparation procedure, and the error on
it was found from a representative measurement. In the later experiments,
instead, vesicles were examined individually, and the error on $\lambda$,
amounting to about 20\%, could be estimated from direct measurements
(see error estimates in \citet{JulienShear}). Other parameters, like
$r_{0}$ and $\Delta$, were calculated from the cross-section measurements
of each vesicle with an error of about 20\% each, as well. Individually,
$\chi$ depends cubically on $r_{0}$ (eq.~(\ref{eq:CaSLambdaShear})),
and thus can vary strongly for different vesicles, even at the same
shear rate in channel flow. The data ensemble was binned and averaged
for some class values of $\Delta$, and was presented as 
$\langle\theta(\Delta)\rangle$
for each $\lambda$ available. For example, each curve in Fig.~1
of \citet{KantslerPRL2006} includes points resulting from many measurements
on vesicles with different $\Delta$ and $\chi$. At that time, from
these ensemble averages, we concluded that $\theta$ does not depend
on $\chi$, in contrast to the recent theoretical findings of \citet{Kaoui}.

\subsection{\label{sub:Tank-Treading}Tank Treading: scaling of
$\theta(\Lambda)$}

A benchmark for the theories is the prediction of the vesicle inclination
angle $\theta$ during TT, as function of the relevant control parameters.
We have addressed the question in our early
papers\cite{KantslerPRL2005,KantslerPRL2006}.
Very recently \citet{Farutin} have analyzed the problem using
a part of our previously published data (from the figures
in \citet{KantslerPRL2006})
and found it in good agreement with their model. Their presentation
can be easily compared with that of a larger set of the data, in similar
variables, shown by \citet{VlakhovskaGracia}.
Moreover, the plot in \citet{Farutin} suggests that all data refers
to $\chi$=100, while in \citet{KantslerPRL2006} the information about
$\chi$ was not provided, for the reason just explained in section
\ref{sub:methods}. To clarify the situation we rediscuss the old
methodology and present more data supporting our view.

\begin{figure}
\noindent \includegraphics[width=1\columnwidth]{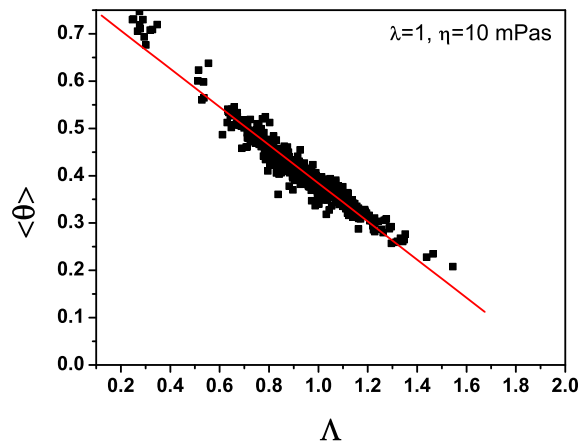}
\caption{\label{fig:Vasiliy1}Inclination angle from ensemble of single vesicles
in TT regime as a function of $\Lambda$ for $\lambda=1$ and $\eta=10$mPa~s.
The straight line is is a linear fit to the data points.}

\end{figure}

\begin{figure}
\noindent \includegraphics[width=1\columnwidth]{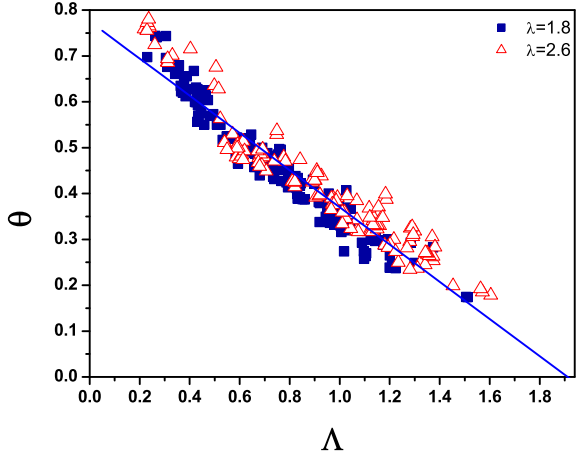}
\caption{\label{fig:Vasiliy2}Inclination angle from an ensemble of individual
vesicles in TT regime as a function of $\Lambda$, for two values
of $\lambda$. The straight line is a linear fit to the data points.}
\end{figure}

The data for $\theta$ can be plotted versus the one or the other
representative quantity, like $\Delta$ or $\Lambda$; a functional
dependence can be sought for single vesicles or for ensemble averages.
In Fig.~\ref{fig:Vasiliy1} and \ref{fig:Vasiliy2} we show $\theta(\Lambda)$
for single vesicles in the TT regime, respectively for $\lambda=1$
and for $\lambda=1.8,\,2.6$. In spite of the scatter of the data
due to thermal noise, the data collapse remarkably when plotted against
the scaled variable $\Lambda$, and correlate well with a linear fit.
The value $\Lambda_{c}$, defined by the intercept of the fit with
$\theta=0$, determines the transition, which is either to TU or to
TR depending on $S$ as discussed further. Compatible $\Lambda_{c}$
are obtained in both plots, within the experimental error bars. The
same was found for old data with $\lambda=1$ and $\eta=1.1$mPa~s
(not shown here). The supplementary material at {[}URL will be inserted
by AIP{]} includes tabular data for $\lambda=1.8$ and $2.6$, with
values of $\chi$ for each vesicle.

\begin{figure}
\noindent \includegraphics[width=1\columnwidth]{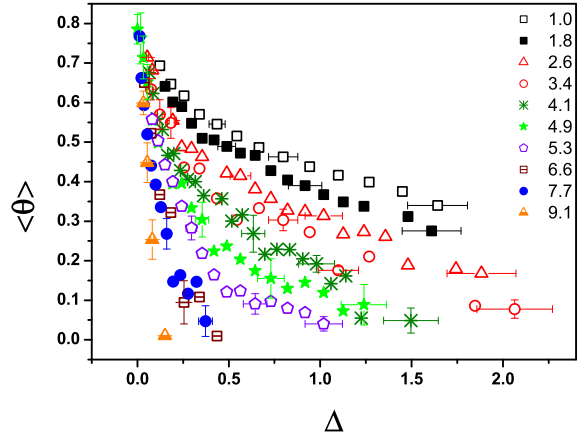}
\caption{\label{fig:Vasiliy3}Mean inclination angle obtained due to ensemble
averaging in TT regime as a function of $\Delta$ for vesicles with
different values $\lambda$ presented on the plot (color online).}
\end{figure}

\begin{figure}
\noindent \includegraphics[width=1\columnwidth]{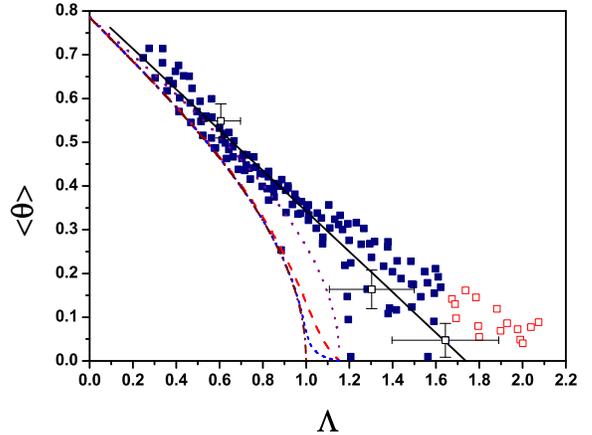}
\caption{\label{fig:Vasiliy4} 
$\langle\theta(\Lambda)\rangle$ as a function
of $\Lambda$, for the data in TT regime
presented in Fig.~\ref{fig:Vasiliy3}, with some typical error bars.
The full as well as the open squares present all the data with 
different $\lambda$ and $\Delta$; the open squares indicate the 
data at small $\theta$ and large $\Delta$ at various $\lambda$, 
susceptible to enhanced thermal fluctuations. 
The dash-dotted line (dark red online) is the theoretical
solution $\theta=\frac{1}{2}\cos^{-1}\Lambda$; the other three
reference lines are the LTV fixed point solution discussed in text,
respectively for $S=50$ (short dash, blue online), $S=10$ 
(long dash, red online), and $S=0$ (dot, purple online).
The full straight line is a linear fit to
the data, based on the full squares only,
$\langle\theta\rangle=0.81-0.46\Lambda$ with $\Lambda_{c}=1.74$.}
\end{figure}

In Figure \ref{fig:Vasiliy3} we plot the data for
$\langle\theta(\Delta)\rangle$
of vesicle ensembles. Part of this data were published in
\citet{KantslerPRL2006},
and are supplemented here by more values of $\lambda$. All available
ensemble averaged data, at different $\lambda$, again collapse when
plotted as function of the scaled variable $\Lambda$ in Fig.~\ref{fig:Vasiliy4}
(analogously to what was done by \citet{VlakhovskaGracia}). The dependence
of $\langle\theta(\Delta)\rangle$ appears as a power law, except
for tails at small $\theta$ and large $\Delta$, seen for instance
in Fig.~\ref{fig:Vasiliy3}, in particular for $\lambda=3.4$, $4.1$,
$4.9$ and $5.3$. The tails are strikingly similar to those observed
in the recent 2D numerical simulations of \citet{Messlinger}
(cfr.~their Fig.~4) at comparable values of $\langle\theta\rangle\leq0.15$
rad and $\Delta\geq0.7$. Their tails are explained by the strong
amplification of thermal fluctuations in the vicinity of the transitions
to either TR or TU. The large scatter of the data at small $\theta$
and large $\Delta$ is responsible, in first instance, for deviations
from the theoretical scaling in Fig.~2 of \citet{VlakhovskaGracia}.
The extrapolated value $\Lambda_{c}$ corresponding to the intercept
$\theta=0$, marking the regime transition, cannot in fact be compared
with a theory that ignores thermal noise. For these tails, the scaling
exponent $\alpha$ in the dependence $\lambda_{c}\sim\Delta^{\alpha}$
(where $\lambda_{c}$ is defined by extrapolation as with $\Lambda_{c})$,
was found in \citet{KantslerPRL2006} to be about $-\frac{1}{4}$ rather
than $-\frac{1}{2}$, as predicted later by the theory and found for
the new data\cite{JulienShear,JulienMill}.

Excluding the data points related to the tails at small $\langle\theta\rangle$,
large $\Delta$ and the values of $\lambda$ mentioned above, the
full set of Fig.~\ref{fig:Vasiliy4} can be fitted within the error
bars by $\langle\theta\rangle=0.81-0.46\Lambda$. This fit provides
$\Lambda_{c}=1.74\pm0.2$. We emphasize that this data was obtained
in shear flow at $\chi>1$, which corresponds, according to
eq.~(\ref{eq:CaSLambdaShear}),
to $S>S_{c}=\sqrt{3}$. In light of what became clear later on about
the phase diagram for vesicles (see section \ref{sec:Phase-diagrams}),
this means that the transition at these values of $S$ is from TT
to TR, and not to TU, as was suggested in \citet{KantslerPRL2006}.

\begin{figure}
\noindent \includegraphics[width=1\columnwidth]{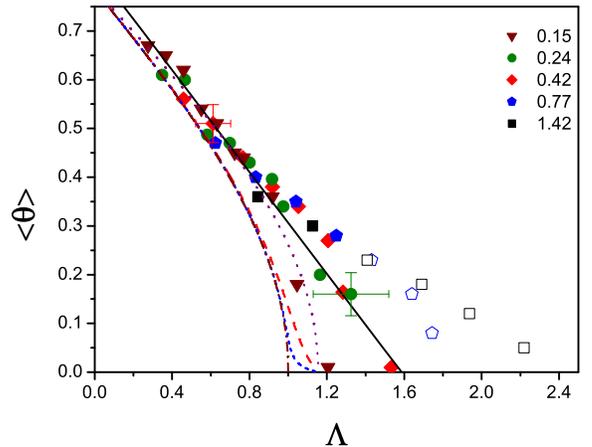}
\caption{\label{fig:Vasiliy5} $\langle\theta(\Lambda)\rangle$ as a function
of $\Lambda$ for the TT  data presented in Fig.~\ref{fig:Vasiliy3},
grouped in classes of $\Delta$. The data points presented by open symbols
corresponding to $\Delta=$0.77 and 1.42 and $\theta\leq 0.2$ are related to
the enhanced thermal fluctuations at small $\theta$ and large $\Delta$. 
The full straight line is a linear
fit to the data, based on the full symbols only,
$\langle\theta\rangle=0.83-0.52\Lambda$
with $\Lambda_{c}\simeq1.6\pm0.2$. Reference dashed lines are the
same as in Fig.~\ref{fig:Vasiliy4} (color online).}
\end{figure}

While there is experimental evidence for scaling of
$\langle\theta(\Lambda)\rangle$,
a theoretical regression law is more elusive. Theories by
\citet{Seifert} for $\lambda=1$, and \citet{Misbah2006} and
\citet{VlakhovskaGracia} for $\lambda\geq1$,
provide an exact solution for the inclination angle in TT up to its
transition to TU, $\theta=\frac{1}{2}\cos^{-1}\Lambda$, giving $\Lambda_{c}=1$.
These theories do not account for a transition from TT to TR, whereas
the LTV theory\cite{LTVnjp,LTVpre} discusses both possible transitions
in detail. According to LTV, the transition TT-to-TU occurs for $S\leq\sqrt{3}$
at $\Lambda_{c}=2/\sqrt{3}\approx1.155$, while the transition TT-to-TR
takes place at $\Lambda_{c}$ up to $\sqrt{2}\approx1.41$. The solution
for the inclination angle in TT is found by solving the system (\ref{eq:DBPVM})
(with $\Lambda_{1}=\Lambda_{2}=0$) with l.h.s.~equal to zero. From
the first of eq.~(\ref{eq:DBPVM}), it can immediately be seen that
this leads to $\theta=\frac{1}{2}\cos^{-1}\left[\Lambda\cos\Theta\right]$,
where $\Theta$ can be represented in terms of $S$ and $\Lambda$.
After some algebra, a closed solution $\theta(\Lambda,S)$ is found,
which is weakly dependent on $S$ in the range $0\le\Lambda<2/\sqrt{3}$,
and bounded from above by its limit for $S=0$, i.e.\
$\theta(\Lambda)=\frac{1}{2}\cos^{-1}\frac{\sqrt{3}\Lambda}{2}$.
This solution is displayed, for reference, for the specific values
$S=10$ and $50$ in Figs.~\ref{fig:Vasiliy4} and \ref{fig:Vasiliy5}.
The LTV solution is closer to the experimental data but still
disagrees with it at $\Lambda\geq 0.8$ and $\langle\theta\rangle\leq0.35$.
Moreover the LTV theory predicts, for $S>\sqrt{3}$ and
$2/\sqrt{3}<\Lambda<\sqrt{2-2/S^{2}}$,
\emph{negative} vesicle inclination angles, which we didn't observe,
and a TT motion which is unstable in the third dimension and thus
in practice not realized. Presenting part of the data of Fig.~\ref{fig:Vasiliy4}
binned in classes of $\Delta$, rather than $\lambda$, we show clearly
in Fig.~\ref{fig:Vasiliy5} that either analytical solution matches
satisfactorily the data only for $\langle\theta\rangle\geq 0.35$ and
$\Delta\leq1.42$. Better agreement between the LTV theory and the
experiment is found only for the data in the lowest $\Delta$ bin,
not surprisingly after the basic assumption of the theory, $\Delta\ll1$.
Thus, the extension of theoretical results beyond $\Delta\approx0.15$
and small $\theta$, used in particular in the recent publications
\cite{Danker,Kaoui,Farutin}, is unreasonable as it violates a basic
assumption.

All experimental points corresponding to $\Delta$=0.15, 0.24, and 0.42 at
all values of $\langle\theta\rangle$ (except for two at $\Delta=0.15$ and
$\langle\theta\rangle$ close to zero) lie on the fitting straight
line  shown in Fig.~\ref{fig:Vasiliy5}, within the error bars.
As for the comparison between the part of the data presented
in Fig.~\ref{fig:Vasiliy5} and the theory of \citet{Farutin},
one cannot distinguish the theoretical
curves corresponding to the lower $\Delta$ within the error bars.
Only the data points shown by the open symbols, corresponding to
$\Delta=0.77$ and $1.42$ and $\theta\leq 0.2$ and related to the enhanced
thermal fluctuations at small $\theta$ and large $\Delta$ deviate from it.
Thus, in our opinion, the quality of the experimental data does not
allow one to distinguish between sets with different $\Delta$ within
the error bars. Besides, as we demonstrated above, the full set of
the data presented in Fig.~\ref{fig:Vasiliy4} is also fitted rather
well by a straight line in the whole range of $\langle\theta\rangle$
(once more, when the data points shown by open squares, related to the
enhanced thermal fluctuations at small $\theta$ and large $\Delta$ at
various $\lambda$ are excluded).
This conclusion brings us back to the problem, discussed in the recent
theoretical\cite{LTVnjp,LTVpre,Danker,Kaoui,Farutin,Vlakhovska}
and experimental papers\cite{JulienShear,JulienMill}, whether a
two or three-dimensional phase diagram is required to present all
vesicle dynamical states. According to our statement, only two parameters
are sufficient to account for the TT data within the error bars. If
\citet{Farutin} were correct, the value of the transition $\Lambda_{c}$
should depend on $\Delta$, while we claim that for what can be understood
from the available data, it is not. Further evidence for the regimes
of TR and TU is discussed below in Section \ref{sec:Phase-diagrams}.

On the other hand, we found surprising and probably accidental that
the linear approximation for $\Lambda\ll1$ to the first solution,
i.e.\ $\theta\approx\pi/4-\Lambda/2$, describes the data rather
well inside the error bars, and would give $\Lambda_{c}=\pi/2\approx1.57$.
This was already pointed out in our early paper\cite{KantslerPRL2005}
and actually even used by \citet{VlakhovskaGracia}. The linearization
of the solution $\theta(\Lambda,S)$ of LTV would have a slightly
milder slope and be as well compatible with the data, with $\Lambda_{c}$
up to $\pi/\sqrt{3}\approx1.81$.

\begin{figure}
\noindent \includegraphics[width=0.9\columnwidth]{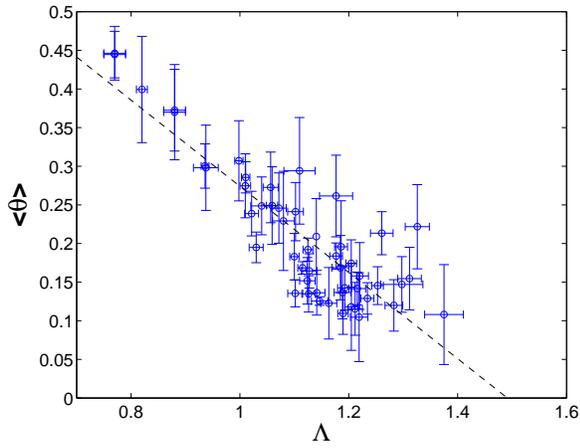}
\caption{\label{fig:TT-Mill-data}$\left\langle \theta\right\rangle $ versus
$\Lambda$ in TT regime for vesicles with $\lambda=1$ in four-roll
mill experiment. The dashed line is a linear fit to the data
 $\left\langle \theta\right\rangle =0.83-0.56\Lambda$
with $\Lambda_{c}=1.49$. Data points with
$\left\langle \theta\right\rangle <0.1$
and standard deviation larger than $0.07$ are removed.}
\end{figure}

The new data\cite{JulienShear,JulienMill} were obtained
differently. The determination of the regime of motion of each vesicle
as a function of $\Lambda$ was conducted as explained in subsection
\ref{sub:methods}. The experiments on vesicle dynamics were conducted
in two different experimental devices and flow configurations. In
the plane Couette flow apparatus only a minority of vesicles happened
to be observed in the close vicinity of the transition, since $\lambda$
and $\Delta$, on which the parameter $\Lambda$ depends, were not
controlled but only measured\cite{JulienShear}. Besides, a much
smaller vesicle population was studied than in the old channel flow
experiments; averaging on $\theta$ was done on short time series
for single vesicles and not for ensembles. The resulting data is too
sparse to analyze the TT motion and scaling, and, in spite of the
smaller error bars and uncertainty in the determination of $\Lambda$
and $S$, the transition lines on the phase diagram in e.g.\ Fig.~6
of \citet{JulienShear} are marked by rather wide bands. In the four-roll
mill device\cite{JulienMill} instead, the control parameter $\omega/s$
was varied for each single vesicle with $\lambda=1$ and measured
$\Delta$, and more data is available. In Fig.~\ref{fig:TT-Mill-data}
we plot the data for $\langle\theta(t)\rangle$, averaged over the
time versus $\Lambda$ obtained in the four-roll mill device with
the latter procedure, for vesicles in TT regime at $S>\sqrt{3}$.
Again, data can be compared to theories only when the noisiest data,
i.e.\ at small $\theta$ and large $\Delta$, is excluded. The data
that is fairly fitted by $\left\langle \theta\right\rangle =0.83-0.56\Lambda$,
giving $\Lambda_{c}=1.49$, which is comparable to the value obtained
for the old data (Fig.~\ref{fig:Vasiliy4} and \ref{fig:Vasiliy5}),
to the approximate solutions, and even to the upper theoretical value
$\Lambda_{c}=\sqrt{2}$ provided by LTV for unstable TT.

In summary, the collapse of all experimental data of
$\langle\theta\left(\Lambda\right)\rangle$
supports only qualitatively the theoretical suggestion of the scaling
\cite{Seifert,Misbah2006,VlakhovskaGracia,LTVnjp,LTVpre}. This analysis
also confirms that neglecting the thermal fluctuations leads to the
scaling $\lambda_{c}\sim\Delta^{-1/2}$, which follows from $\Lambda_{c}=const$,
for both the old and new experimental data.

\subsection{Trembling: Analysis of experimental vesicle contours, dynamics of
harmonics and thermal noise\label{sub:ExperimentModes}}

\begin{figure}
\begin{centering}
\includegraphics[width=0.3\columnwidth]{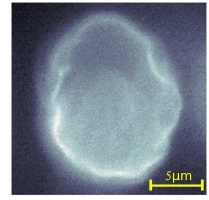}
\par\end{centering}
\caption{Image of a vesicle in TR, exhibiting multiple out-of-focus indentations
suggesting a three-dimensional wrinkling-like perturbation. Quantitative
analysis of the instantaneous three-dimensional shape is not possible
with our technique.\label{fig:3d-TR-image}}
\end{figure}

\begin{figure*}
\begin{centering}
\includegraphics[width=0.7\textwidth]{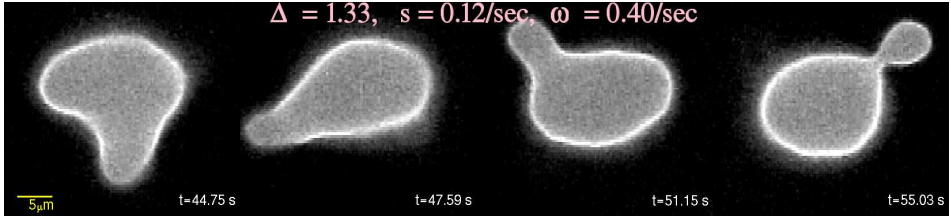}
\par\end{centering}

\caption{Vesicle with $\lambda=1$, $\Lambda=2.71$ and $S=2.43$, generating
a protrusion (pinching). The movie is available in the Supplementary
Material {[}URL will be inserted by AIP{]}.\label{fig:outbudding}}

\end{figure*}

\begin{figure*}
\begin{centering}
\includegraphics[width=0.7\textwidth]{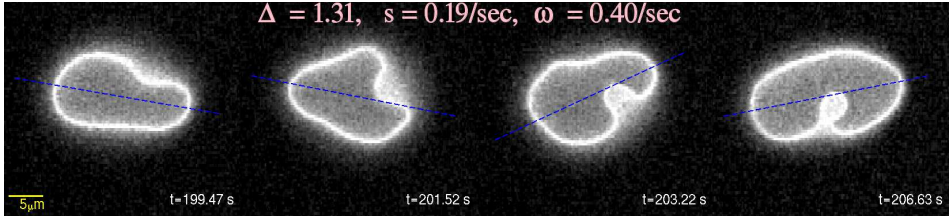}
\par\end{centering}

\caption{Trembling vesicle with $\lambda=1$, $\Lambda=1.72$ and $S=9.75$,
budding. The movie is available in the Supplementary Material {[}URL
will be inserted by AIP{]}.\label{fig:inbudding}}

\end{figure*}

TR, which is the intermediate state between TT and TU, turns out to
be the key regime to understand the vesicle dynamics in a general
linear flow. In this regime, the inclination angle $\theta$ oscillates around
zero. During an oscillation cycle a given membrane patch periodically
experiences both stretching and compression. Because of that, the
TR dynamics is found to be more complex than even TU. The latter is
also characterized by the periodic switching between stretching and
compression, but the time spent under compression at small inclination
angles in TR is much longer. This circumstance leads to stronger vesicle
shape deformations in the TR regime due to the volume and surface
area constraints and to extreme sensitivity to thermal noise at small
$\theta$. The occurrence of strong shape perturbations and the appearance
of higher order harmonics resemble very much the wrinkling recently
observed and studied in a time-periodic elongation flow
\cite{expwrinkling,theorywrinkling}.
In the latter case, the control parameter, which is the elongation
rate, could be varied in order to find the onset of the instability
and to study the nonlinear dynamics of higher order modes, above the
onset. Here, like in TR, higher order modes (wrinkles) are generated
during the compression period, due to the constraints. During compression,
strong shape deformations present mostly as concavities of the vesicle,
producing locally a negative surface tension, which in turn initiates
the instability, with strong sensitivity to thermal noise. As the
result, both even and odd higher order harmonics are generated. Their
growth is arrested since compression acts for just a brief part of
the period, but sometimes, at larger noise amplitude, vesicle budding
occurs\cite{expwrinkling}. Similar effects of vesicle wrinkling
and budding and pinching have been occasionally observed also in TR
(see Figures \ref{fig:3d-TR-image},~\ref{fig:outbudding} and \ref{fig:inbudding}
and movies {[}URL will be inserted by AIP{]}). However, even the snapshots
of a vesicle performing a more regular TR (e.g.~Fig.~\ref{fig:Imagery})
clearly demonstrate a drastic difference with the snapshots
of the vacillating-breathing mode presented in \citet{Danker}.
In the VB mode the vesicle shape remains elliptical, within only even
second and fourth harmonics and a vesicle indeed imitates breathing.

To study quantitatively the TR dynamics, we analyzed the images of
the experiments \cite{JulienShear,JulienMill}, looking at the radial
amplitude of the contour $r(\phi,t)$, $0\le\phi\le2\pi$, relative
to the centroid of the vesicle contour in each image. This radial
profile is Fourier-decomposed, i.e.\ $r$ is expressed as
$r(\phi,t)=\sum_{q}\tilde{r}_{q}(t)e^{iq\phi}$.
Observations about the interplay of different modes in time are presented
in the following. We observe as a side remark, that even performing
the Fourier decomposition in the frame of reference centered on the
vesicle, there is no a priori geometrical symmetry guaranteeing that
any particular contour harmonic is null or conserved. A translation
of the frame of reference would alter all modes of the decomposition,
but the fact that a particular mode (e.g.\ $q=1$) is not null, is
not at all an indication of a miscentered frame of reference.

\begin{figure}
\noindent \begin{centering}
\includegraphics[width=0.99\columnwidth]{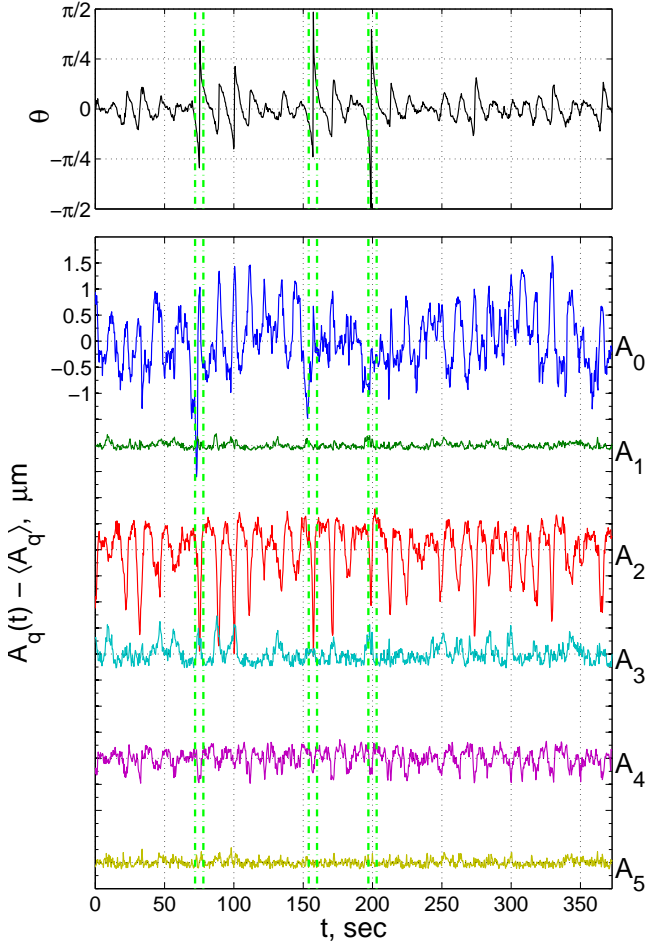}
\par\end{centering}

\caption{\label{fig:harmonic-timeline}Upper panel: inclination angle
$\theta(t)$ of of a vesicle with $\Delta=1.16$ and $\lambda=1$, during
the course of a long trembling
sequence. Lower panel: amplitudes $A_{q}(t)$ of the first harmonic modes
of the sectional contour. For convenience of plotting, mean values
have been subtracted, and time traces have been shifted vertically
relative one to another $2\mu m$. Vertical dotted lines are added
as a guide to the eye, to mark events of occasional\emph{tumbling}
(color online).}

\end{figure}

\begin{figure}
\noindent \begin{centering}
\includegraphics[width=0.99\columnwidth]{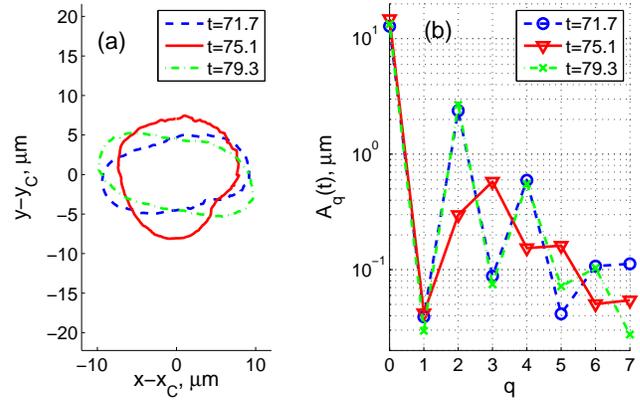}
\par\end{centering}

\caption{\label{fig:spectra-and-contours}
(a) Instantaneous contours in center-of-area
coordinates, and (b) angular spectra of selected snapshots from the
trembling sequence of figure \ref{fig:harmonic-timeline}.
Note the significant decrease of $A_{2}$ and $A_{4}$ and the corresponding
increase of $A_{3}$ for the intermediate contour (color online).}

\end{figure}

Figure \ref{fig:harmonic-timeline} displays the time evolution of
the inclination angle $\theta(t)$ and of the amplitude of the lowest
harmonic modes $A_{q}(t)=\left|\tilde{r}_{q}(t)\right|$ for a typical
long trembling sequence, at constant $s$ and $\omega$. Trembling
motion is seen to be characterized by recurrent, roughly periodic
oscillations in the amplitude of the zero-th mode, accompanied by
short duration dips in the amplitude of the second mode, correlated
with peaks of the third. The latter means that the vesicle section
periodically departs for a short time from its oval shape, to attain
a more triangular appearance. This activity reflects fluctuations
of the three-dimensional vesicle shape, since the fluctuations in
the observed contour length corresponding to the perturbations in
$A_{0}$ can occur only due to 3D effects, which are indeed observed
in the experiment and resemble wrinkling (see Fig. \ref{fig:3d-TR-image}).
Some instantaneous contours of the vesicle during this sequence are
shown in Figure \ref{fig:spectra-and-contours} together with their
spectra, notably around the time of one of such deformations. The
intermediate power spectra in Fig.~\ref{fig:spectra-and-contours}b
clearly demonstrates the prevalence of the third mode over fourth
and even second modes at the time when the vesicle contour
(see Fig.~\ref{fig:spectra-and-contours}a)
is sort of triangular with an additional concavity. This is apparent
in the movie provided as Supplementary Material {[}URL will be inserted
by AIP{]}. From Figure \ref{fig:harmonic-timeline} we also see that
the time traces of the first few harmonics are correlated with modes
2 and 3, and that higher harmonics are decreasingly smaller and noisier
as $q$ increases. The inclination angle $\theta$ is seen to oscillate
somewhat irregularly around zero, while the vesicle occasionally performs
full tumblings. We observe this motion notably in flow conditions
close to the TR/TU transition (see section \ref{sec:Phase-diagrams}
below). Such irregularities led us to use the simple and robust regime
classification criterion mentioned above.

\begin{figure*}
\noindent \centering{}\includegraphics[width=0.6\textwidth]{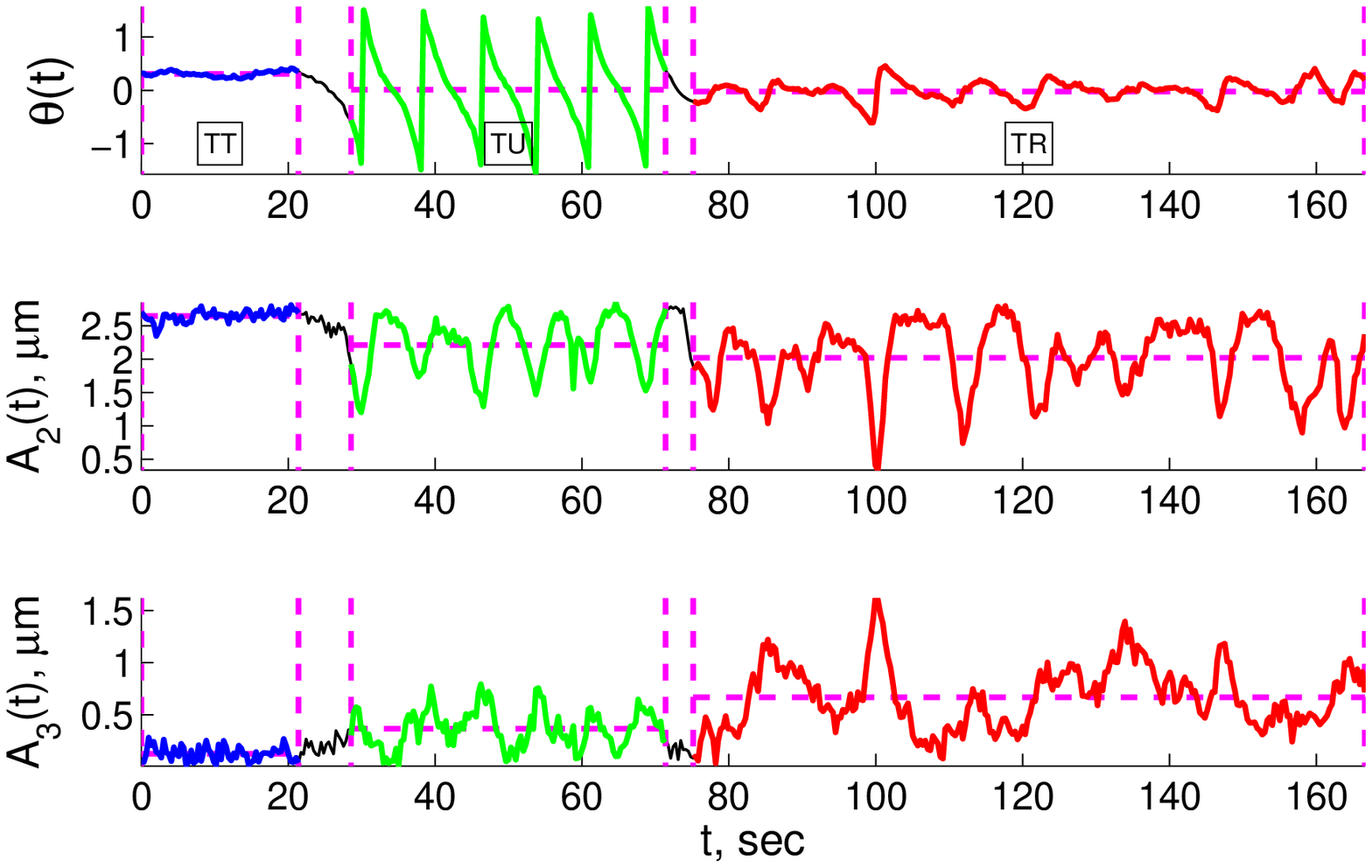}
\caption{\label{fig:RegimeTransient}Instantaneous inclination angle $\theta(t)$
and amplitudes of the second and third angular harmonics $A_{2}(t)$
and $A_{3}(t)$ for a vesicle with $\Delta=0.71$ and $\lambda=1$ in a
general flow,
with $s$ and $\omega$ kept constant for extended time and then quickly
changed. Vertical (magenta online) dashed lines delimit stretches
of constant flow, while horizontal dashed lines mark time averages.
Note that the transition between one regime of motion and the next
is sudden, and the transient appears to be shorter than one period
(of either TR or TU) (color online).}

\end{figure*}

Figure \ref{fig:RegimeTransient} provides an example of a vesicle
driven into the three regimes, evidencing the evolution of the inclination
angle and of the second and third contour modes. The second harmonic
mode $A_{2}(t)$ peaks and dips irregularly during the TR oscillations,
with a correlation of the dips with the minima of the inclination
angle $\theta$; during TU, this mode has minima in correspondence
with $\theta\sim(2n+1)\pi/2$, that is twice per full vesicle revolution.
The increase in $A_{3}(t)$ passing from TT to TU and even more to
TR is apparent, as well as the irregularity of TR motion.

\begin{figure*}
\noindent \begin{centering}
\includegraphics[width=0.7\textwidth]{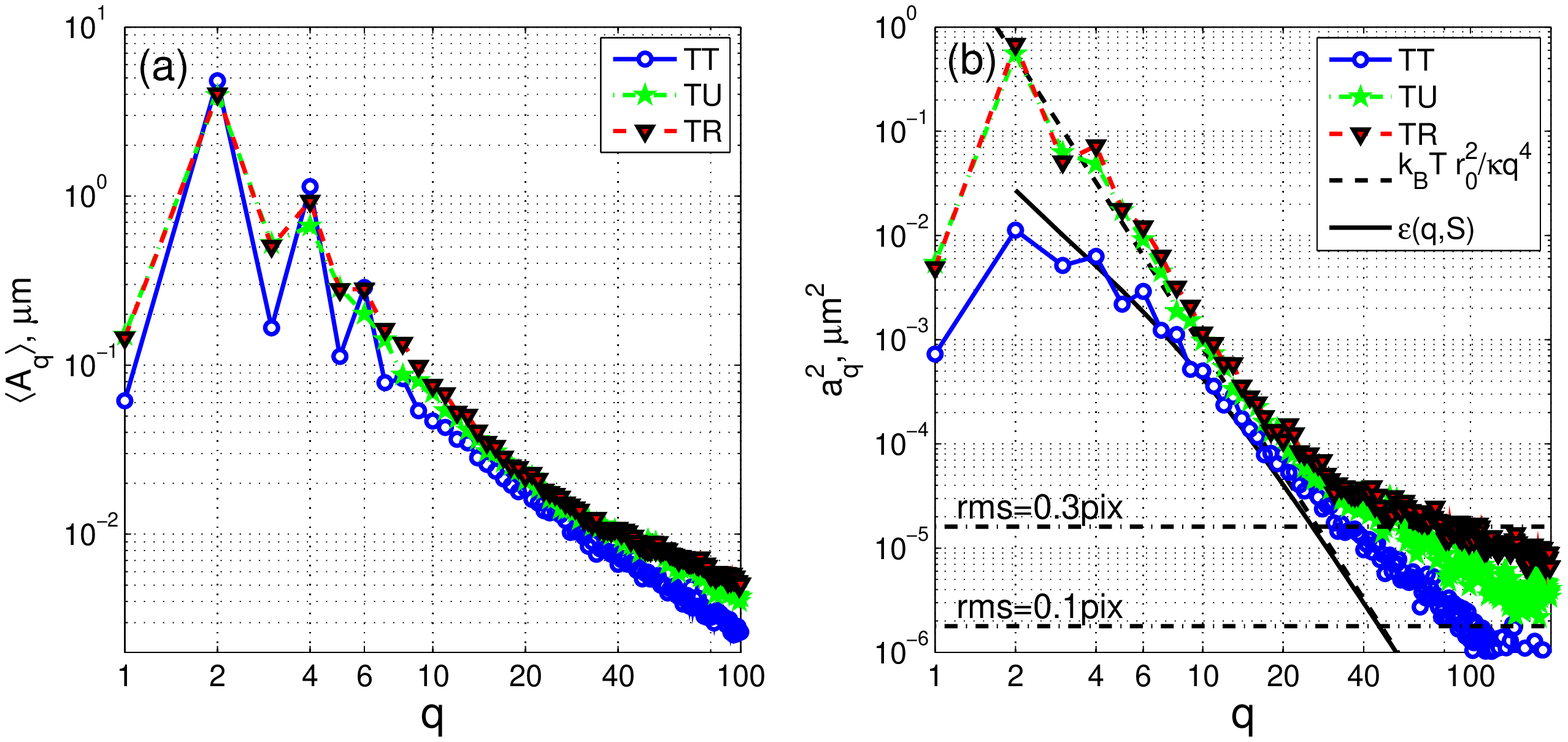}
\par\end{centering}

\caption{\label{fig:harmonic-spectra}(a) Spectra of the mean contour harmonic
modes $\langle A_{q}\rangle$ for a typical vesicle with $\Delta=0.64$ and $\lambda=1$,
accessing the three regimes; (b) spectra of the squared fluctuations
 $a_q^2$ for the same sequence. The solid black line is the theoretical
 form $\epsilon(q,S)$ given in eq.\ (\ref{eq:thermalfluct}), evaluated for
$S=114.11$, corresponding to the TT case. The dashed black line is the limiting
form $\frac{k_B T r_0^2}{\kappa q^4}$. At high $q$, the spectra flatten due to
the contour reconstruction noise. Horizontal dash-dotted lines corresponding
to a white noise of amplitude $0.3$ and $0.1$ pixels are reported
for reference, and demonstrate the subpixel accuracy of our analysis
(color online).}
\end{figure*}

Finally, in Figure \ref{fig:harmonic-spectra}(a) we show the spectra
of the amplitudes $A_{q}=\left\langle A_{q}(t)\right\rangle $ of
the harmonic modes of the contour averaged in time, for another vesicle
with $\Delta=0.64$, which was also driven in the three different
regimes for some time. These spectra are typical of all cases observed,
in the following sense: mode 1 is always small; modes 2 and 4 have
mean amplitudes roughly independent from the regime; mode 3 and 5
are always a few times smaller in TT than in TR or TU. The amplitude
$A_{3}$ is smaller for vesicles with smaller values of $\Delta$.
These plots quantify the observation, seen by eye from images like
those in Figure \ref{fig:Imagery}, that vesicles are {}``almost
elliptical'' during TT, while they undergo more elaborate shape changes
during TR and TU. There seems in general to be no dependence, either
on the regime or on $\Delta$, of the decay spectrum of higher harmonics
in the range $q\sim6\div30$.
For the same sequence, the mean squared fluctuations of the mode amplitudes,
$a_q^2=\langle\left(A_q(t)-\langle A_q\rangle\right)^2\rangle$, are shown in
Figure \ref{fig:harmonic-spectra}(b). It is interesting to note that the
fluctuations of the lower order modes are much smaller in TT than in both
TR and TU. The spectrum of fluctuations of a vesicle  can be compared with
the theoretical prediction of \citet{Seifert97} in thermal equilibrium
\begin{equation}
  a_q^2=\epsilon(q,S)=
  \frac{r_0^2 k_B T}{\kappa(q-1)(q+2)[q(q+1)+\sigma r_0^2/\kappa]},
\label{eq:thermalfluct}
\end{equation}
where $\sigma$ is the surface tension. For large $S$ the last term in the
denominator can be expressed\cite{Seifert}
as $\sigma r_0^2/\kappa=0.77S\sqrt{\Delta}$.

For large $q$, the expression (\ref{eq:thermalfluct}) can be rewritten as
\begin{equation}
  a_q^2\simeq \frac{r_0^2 k_B T}{\sigma r_0^2 q^2+\kappa q^4}.
\end{equation}
It is remarkable
in Figure \ref{fig:harmonic-spectra}(b) that the fluctuations for TR and TU
are quite well described by this latter expression, with $\sigma\to 0$. An
explanation of this fact may lie in the fact that the effective surface
tension, which becomes locally negative during the compression of the
membrane, averages out over an oscillation cycle of the vesicle.

\subsection{Regime transitions and oscillation periods of TR and TU
\label{sub:periods}}

Figure \ref{fig:RegimeTransient}, as well as the equivalent Fig.~5
of \citet{JulienShear} and Fig.~4 of \citet{JulienMill}, also demonstrate
the effect of a quick change of the control parameters, $\dot{\gamma}$
in the former and $\omega/s$ and $s$ in the latter,
on the vesicle motion. The time series of $\theta(t)$ shows an almost
instantaneous change of the dynamical regime, within about a period,
much shorter than the characteristic dynamical time $\tau$. The theory
\cite{Danker} and the associated numerical simulations, without thermal
noise, show instead very long transients, of duration of the order
of $\chi^{-1}$, which are not found in the experiments.

\begin{figure*}
\noindent \centering{}\includegraphics[width=0.6\textwidth]{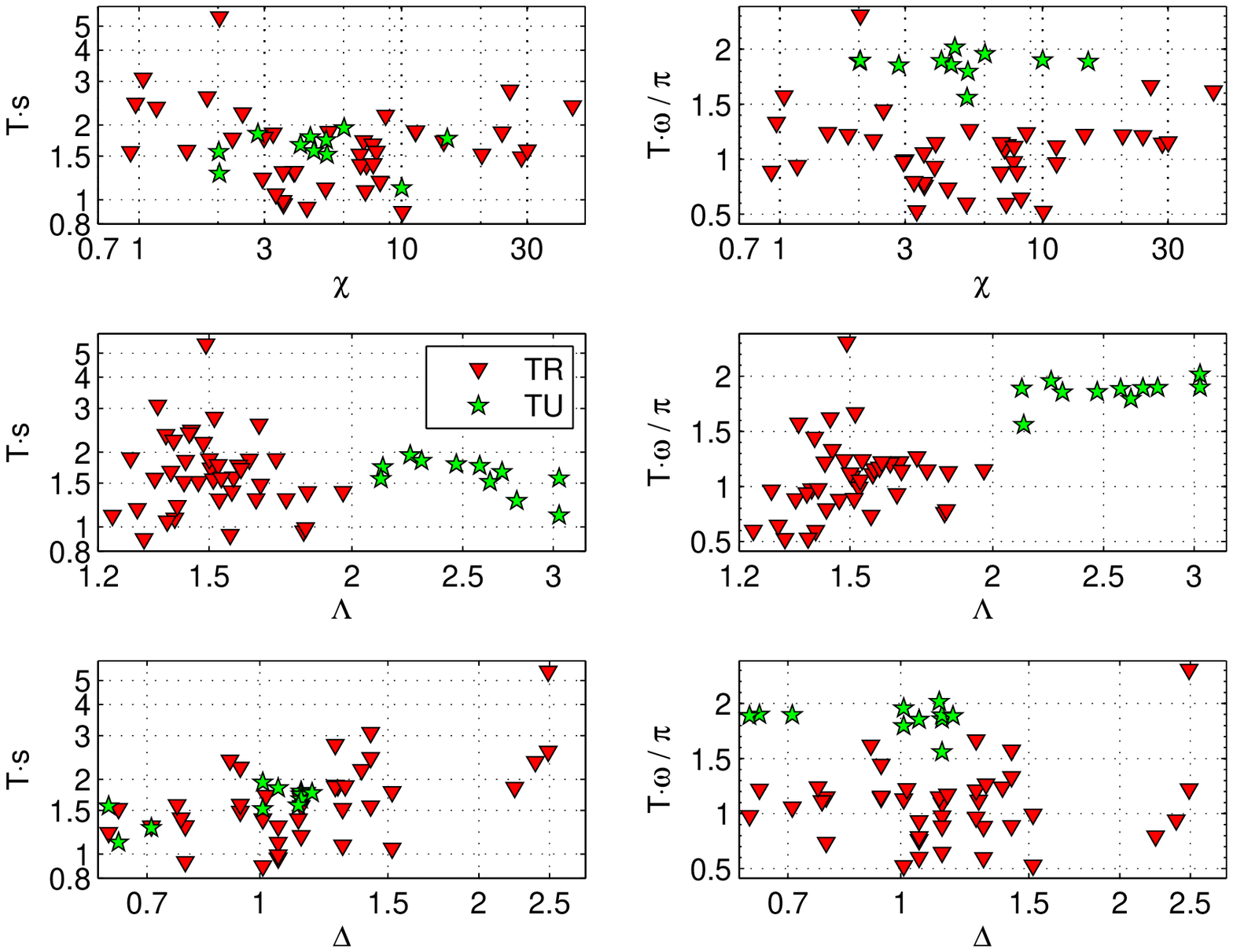}
\caption{\label{fig:Periods}TR and TU periods, rescaled with $s$ and $\omega$
(color online).}
\end{figure*}

The characteristic oscillation period in either TR or TU is extracted
from all available data like that presented in
Figures \ref{fig:harmonic-timeline}
and \ref{fig:RegimeTransient}. To this extent, we determined the
dominant temporal frequency in $A_{2}(t)$ using the \texttt{pburg}
function of \texttt{Matlab}; this frequency is divided by two in TU,
to account for the observed periodicity within one tumbling cycle.
The resulting periods $T$, for the vesicle sequences of duration
sufficient for a reliable estimate, are shown in Fig.~\ref{fig:Periods},
rescaled by either $s$ or $\omega$. Surprisingly, we didn't find
any correlation between $T$ and $\tau$ provided by eq.~(\ref{eq:SLambdaTau}).
We observed a large dispersion of periods in TR, while for large $\Lambda$
the periods tend to $T\simeq2\pi/\omega$ in TU regime, as expected
for rigid body rotation. Correlation with $\Delta$ or $\chi$, as
predicted by theories (e.g.\ \citet[]{Kaoui}) is not observed either.

Both the the fast regime adjustments and the observed oscillation
periods in TU and TR disprove, even qualitatively, the picture presented
by the reduced theoretical models. Instead of a \textquotedbl{}breathing\textquotedbl{}
of the vesicle shape in TR, observed in the models which employ the
second and eventually the fourth order harmonics and without thermal
fluctuations, we observe a noisy dynamics. The experimental picture
found is one of strong mode interaction and correlation, with a pronounced
role of the third harmonic, where thermal noise is considerably amplified.
This underlying mechanism is very different from the viewpoint of the papers of Misbah's group
\cite{Misbah2006,Danker,Kaoui,Farutin}, though the vesicle shape
deformations are comparable with those presented by \citet{Messlinger}.
For better appreciation the movie of TR dynamics provided in Supplementary
Material {[}URL will be inserted by AIP{]} should be compared with
the snapshots of the VB mode presented in Fig. 5 of \citet{Danker}.

\section{Phase diagrams - Comparison with recent experimental data\label{sec:Phase-diagrams}}

The study of the dynamical systems described in Section \ref{sec:Dynamical-models}
leads to the construction of phase diagrams. That is, regions in the
model parameters space, where the same regime of motion is attained.
A similar approach has been applied to capsules \cite{KesslerFinkenSeifert,KesslerFinkenSeifert-swinging,SkotheimSecomb,Bagchi}.
While all authors agree on the existence of at least the three regimes
of motion mentioned above, they dissent about the dimension of the
phase space, and about the position of the regime boundaries in that
space. According to LTV, the phase space is $\{S,\Lambda\}\in\left[0,\infty\right[\times\left[0,\infty\right[$;
according to DBPVM and KFM the phase space is three dimensional, and
better represented by the group $\{Ca,\lambda,\Delta\}$. The choice
of LTV scaling has clearly the advantage of simplicity, and, as we
want to demonstrate, probably accounts for the correct scaling in
powers of $\sqrt{\Delta}$, though it agrees only qualitatively with
the experiment. To assess this, we plot all available data in the
one or the other parameter space, and compare the results.

\begin{figure}
\noindent \includegraphics[width=1\columnwidth]{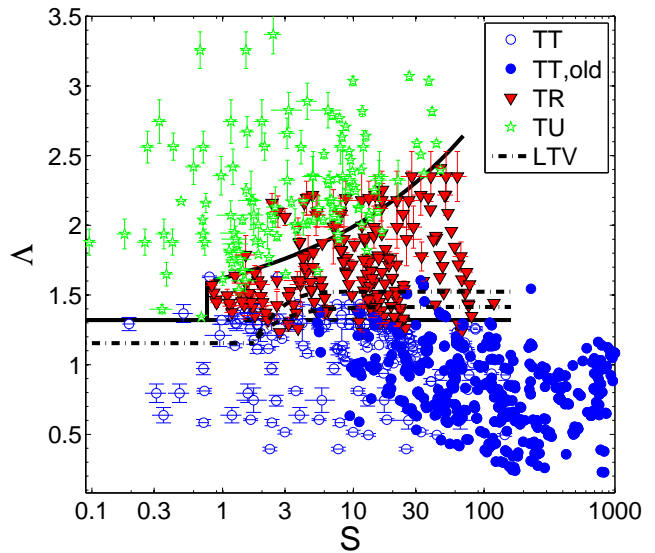}
\caption{\label{fig:phase-diagrams-SL}Summary of the motions observed, with
coding based on the regime -- TU, green stars; TR, red triangles,
and TT, blue circles (full symbols - older data\cite{KantslerPRL2005,KantslerPRL2006};
empty symbols - newer data\cite{JulienShear,JulienMill}).
Data points are plotted with error bars accounting for the measurement
uncertainties. The dash-dotted lines are the dividers given by \citet{LTVnjp},
while the solid lines are a guide to the eye (for details see
text) (color online).}

\end{figure}

Figure \ref{fig:phase-diagrams-SL} shows our most recent phase diagram,
which includes data for vesicles with $0.015<\Delta<2.5$ from the
experiments\cite{KantslerPRL2005,KantslerPRL2006,JulienShear,JulienMill}
and extends the parameter ranges. Classification of regimes is automatic
for the data of \citet{JulienMill} and obtained from the time series
of the vesicle inclination angle, as explained in Section \ref{sub:ExperimentModes}.
Data from \citet{JulienShear} had larger error bars and the regime
classification was done by eye; older data\cite{KantslerPRL2005,KantslerPRL2006}
is only TT (see supplementary material at {[}URL will be inserted
by AIP{]} for all data in tabular form). With this caveat and furthermore,
with the ambiguities of regime identification of deformed vesicles
close to the transition lines, we claim that clustering in different
regions is clear. We plot, for reference, the dividers of the regimes
given by \citet{LTVnjp}, respectively
 $\left\{ S<\sqrt{3},\,\Lambda=2/\sqrt{3}\right\} $
for TT/TU, $\left\{ S\ge\sqrt{3},\,\Lambda=\sqrt{2-2/S^{2}}\right\} $
for TT/TR, and
 $\left\{ S\ge\sqrt{3},\,\Lambda=1.52-2.12\, e^{-1.04\, S}\right\} $
for TR/TU (best fit based on the numerical data of their Figure 9).

The extent of the experimental regions deviates from the model. The
solid lines in the figure are our eyeball fit to
the data, represented by: $\{0<S<0.75,\,\Lambda=1.32\}$
(TT/TU and TT/TR divider) and
$\left\{ S>0.75,\,\Lambda=1.32+0.3\, S^{0.35}\right\} $
(TR/TU divider). Note the lower $\Lambda$ transition
(to tank treading) has zero slope on both diagrams in the range of
our data, for all $\Delta$ sampled. The upper line has finite slope
for large $S$, and disagrees with the results of all reduced model
simulations. The divider between TR and TU is also less defined on
the diagram, in part due to some experimental points with large error
bars, but mostly because of ambiguities, namely of vesicles which
intermittently flip between large-amplitude trembling and full tumbling
rotations, as exemplified by Fig.\ \ref{fig:harmonic-timeline},
obtained for $S=4.7\pm0.3$ and $\Lambda=1.59\pm0.03$. A possible
explanation for the fact that the upper transition differs greatly
from the linear theories and the lower transition does not, is that
the former is associated with a saddle-node bifurcation, while the
latter with a Hopf type one. It is well known that the nonlinear dynamics
associated with the saddle-node transitions are more sensitive to
noise than those near Hopf transitions.

\begin{figure}
\noindent \includegraphics[width=1\columnwidth]{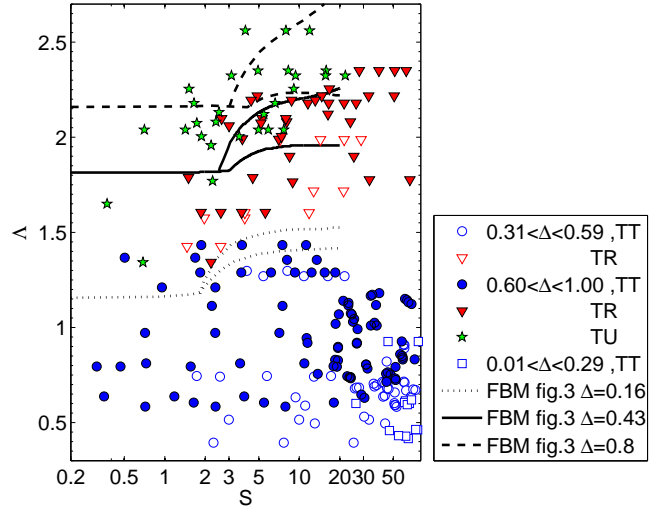}
\caption{\label{fig:phase-diagram-Farutin}Comparison of our experimental results
on the phase diagram with the theory of \citet{Farutin} (color online).}
\end{figure}

The clustering of regimes would be destroyed if the data was plotted,
in coordinates $\{\chi,\lambda\}$. In particular, all the experimental
data for vesicles with $\lambda=1$, any regime of motion, would collapse
on a single horizontal line. According to \citet{Danker,Kaoui}, this
would happen since the correct dependence on the third parameter,
$\Delta$, is not taken into account. However, contrasting this view
we plot a subset of our data taken in pure shear flow, in
Fig.\ \ref{fig:phase-diagram-Farutin}.
To compare with Fig.~3 of \citet{Farutin}, we reproduce their transition
curves and select among the available data vesicles with $\Delta$
close to the values reported by them. It is rather obvious that the
theory\cite{Farutin} disagrees with the data even qualitatively.
We do not attempt comparisons between our data in general flow and
the Fig.~4 and 5 of \citet{Farutin}, as not enough data points
close to the values displayed there are available.

\begin{figure}
\noindent \includegraphics[width=1\columnwidth]{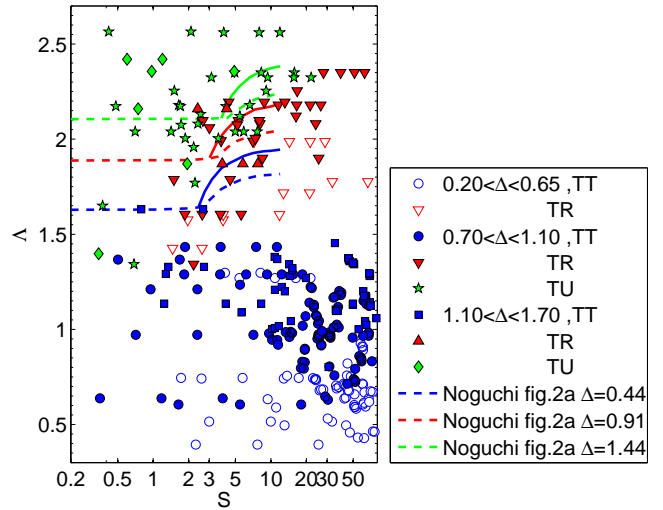}
\caption{\label{fig:phase-diagrams-Noguchi}Comparison of our experimental
results on the phase diagram with the model of
\citet{NoguchiOscillatory} (color online).}

\end{figure}

The dependence on $\Delta$ suggested by \citet{NoguchiOscillatory}
(his Fig.~2a) for vesicles in uniform shear flow is not matched
either by the experimental data, as we show in
Fig.~\ref{fig:phase-diagrams-Noguchi},
though the discrepancy is smaller than with the theory of \citet{Farutin}.
The disagreement between this model, which includes
higher order terms, and the phenomenological model of
\citet{NoguchiOscillatory},
is also apparent when comparing Figures~\ref{fig:phase-diagram-Farutin}
and~\ref{fig:phase-diagrams-Noguchi}.

In summary, we conclude that the coordinates of the LTV phase diagram
($\Lambda$ vs $S$) look preferable due to its simplicity, failing
the quantitative agreement with data. As we already pointed out
\cite{JulienShear,JulienMill}, the LTV theory is based on some assumptions
which are not matched by experimental observations. However, puzzling
as it may be, the presentation of a two-parameter phase diagram resulting
from the self-similar solution of the model provides an adequate description
of the data.

\section{Conclusions\label{sec:Conclusions}}

In attempt to clarify the mechanisms responsible for vesicle dynamics,
we have carefully reexamined existing experimental data, and critically
reviewed the recent theoretical and numerical work. We found that
despite some qualitative features captured by the existing reduced
models, a good quantitative prediction is not achieved. This lack
of success may be due to the incomplete understanding and modeling
of thermal fluctuations, nonlinear interaction of harmonic modes beyond
the second (which can form local regions of negative curvature - which
we have seen in the experiments). Present models are derived in the
nearly-spherical approximation, and even $\Delta\sim0.43$ may be
considered large. The main conclusion of our analysis is that the
agreement between theory and the experimental observations may not
improve, not even on a qualitative level, by including in the existing
models just some of the necessary elements. In particular, including
more even modes without including also odd ones and thermal fluctuations,
has been proven unsuccessful either for the description of the TR
dynamics or the phase diagram presentation. In any event, we hope
that all future modelers will look critically at the effects of noise
on transition regimes where $\langle\theta\rangle<0.15$ radian and where odd
modes of vesicle contours are significant.

\section*{Acknowledgments}

One of us (VS) is grateful to Dr.\ Y.\ Burnishev for his help in
data analysis. This work is partially supported by grants from Israel
Science Foundation, the Minerva Foundation, and by the Minerva Center
for Nonlinear Physics of Complex Systems.


%

\end{document}